# A Hierarchical and Modular Radio Resource Management Architecture for 5G and Beyond

Marcin Dryjanski (IEEE SM), Grandmetric, Poland
Adrian Kliks (IEEE SM), Poznan University of Technology, Poland

*The evolution of mobile wireless systems into Heterogeneous Networks, along with the introduction of the 5th Generation (5G) systems, significantly increased the complexity of the radio resource management. The current mobile networks consist of multitude of spectrum bands, use cases, system features, licensing schemes, radio technologies and network layers. Additionally, the traffic demand is uneven in terms of spatial and temporal domains, calling for a dynamic approach to radio resource allocation. To cope with these complexities, a generic and adaptive scheme is required for efficient operation of Heterogeneous Networks. This article proposes to use a hierarchical and modular framework as an approach to cover the mentioned challenges and to generalize this scheme to different network layers. The proposed management solution is based on three main components: specialized solutions for individual requirements, exposed to the coordination layer, through abstraction middleware. In this approach, new items can be added as "plugins".[1]*

## I. Introduction

Recently, mobile networks evolved into Heterogeneous Networks (HetNets), to improve the capacity in traffic hotspots. The mobile network's landscape gets more complicated due to continuous introduction of new system features, spectrum bands, use cases, network layers (e.g. with the introduction of Small Cells) etc. Spectrum availability, spectral efficiency improvements and higher densification of the radio networks are considered essential with the advent of 5G systems. The advancements within the 3GPP LTE, under the name LTE-Advanced Pro [1], equips it with set of functions like, Dual Connectivity (DC), Massive Carrier Aggregation (CA), Full-Dimension MIMO, or the aggregation of licensed and unlicensed spectrum under the LTE-WLAN Aggregation (LWA). The resources from licensed and unlicensed spectrum can also be aggregated, within the CA framework under the Licensed-Assisted Access (LAA) scheme. Narrowband Internet-of-Things (NB-IoT) addresses the need for low-power Machine-Type Communications (MTC), Device-to-Device (D2D) feature enables direct communication between UEs, while Vehicular-to-Anything (V2X) builds upon D2D framework focusing on the vehicular communication use cases. In parallel to the LTE developments, the work on the 5G systems started already within 3GPP Release 14 [2]. The 5G work was split into Non-Standalone 5G mode and Standalone mode, where the new radio interface provides the signaling connection itself.

5G targets the three key usage scenarios, namely the enhanced Mobile Broadband (eMBB), massive Machine-Type Communications (mMTC), and Ultra-Reliable and Low-Latency Communications (URLLC). The 5G air interface (called New Radio, NR) is designed be able to cope with these use cases and their divergent requirements on high throughput, low latency or massive sporadic transmissions. On top of this, the NR supports the large frequency range, from 450 MHz up to 52.6 GHz (and above as of the current discussions within Rel-17). To capture the differences in propagation and use case requirements, scalable numerology is used in NR. On the network side, the concept of slicing was introduced to capture the differences in the various use cases and to enable optimization of independent logical networks, built upon a single infrastructure. All of the above, provide a landscape of the mobile networks, which exposes the overall high complexity of the system. This particularly touches the area of Radio Resources Management (RRM) and network management. Whenever a new item is added, the RRM complexity is increased. A native and unified approach to the coordination of radio access mechanisms and resources for efficient data delivery is therefore needed. The paper is arranged in the following way: Chapter II provides background for the work and states the problem. Chapter III proposes a design of resource management framework. Chapter IV shows examples of the proposed approach in three use cases, the Unified Medium Access Control (MAC) design, Unified Flow Control and Unified Traffic Steering (UTS). Chapter V formulates a generalization of the framework by proposing extension directions towards other network layers. Finally, Chapter VI summarizes the work.

## II. Overview of Wireless Systems Complexity

### Evolution of Wireless Networks

Looking at the evolution of mobile wireless systems, we can emphasize two major aspects: first, key services, which those systems were designed to provide; second, a general assumption captured when designing them. In that context, authors would like to underline the following. First, the odd generations provide an approach to new requirements, along with services that were not present before (e.g., 1G for mobile voice, 3G for mobile Internet access), while the even generations represent the evolution of the odd ones, improving the design of the predecessors to deliver the main service. Second, until 4G, the network was designed with a "single" design approach, also known as "one-size-fits-all", because there was a single main service to address (e.g. voice or Internet access). With the emerging 5G the perspective has changed. In current situation, there is no single "killer application", to be served by the system, but rather the system should be able to cover multitude of services with diverging requirements. Those services, like high throughput mobile broadband (addressed by eMBB), through extremely low latency communications (with URLLC), down to low-end MTC applications with very sporadic transmissions using several bytes (with mMTC), call for support of different connectivity approaches. The 5G system is designed to be flexible enough to cover those requirements [2]. 5G is the "odd" version, targeting the collection of services for the first

---

time. Whether the approach for 6G will be to "fix the 5G legacy mistakes" – time will tell – but without a new way for the system design, it will be difficult to proceed, because 6G will focus on further extending the service diversification [3].

*Systems Complexity Overview*

The increasing amount of available spectrum resources, equipped with a number of the spectrum access technologies, provides a complicated system to operate, with a complex coordination of the resources, among RAN nodes. An example of a HetNet comprise of macro- and SC layers, accompanied by various spectrum access techniques, including CA and DC, and utilizing both the licensed and unlicensed spectrum bands [4]. On top of this, when moving towards the 5G era, different requirements have been posed by vertical services. However, it is worth mentioning that the system is not supposed to always fulfill all of them at once, but rather deal with specific requirements of the specific service, wherever and whenever needed. In other words, different services require different types of optimizations, e.g. eMBB focuses on high throughput and capacity, while mMTC focuses on long battery life and deep coverage, and URLLC on high reliability and low latency [4]. The evolution of the systems with the growing number of features is shown in *Figure 1*. When considering the complexity of mobile systems, along with their evolution, three main points have been identified:

- network densification techniques, new spectrum access schemes and the growing availability of radio access solutions are all increasing the flexibility of the RRM scope, which yield an increase in its complexity [5];
- considering the variety of the 5G requirements, the RRM scene is highly fragmented and, to fulfil 5G expectations, it is required to provide a unified and scalable approach to radio resources handling [5];
- backwards compatibility, which is a typical requirement for most systems, makes the ecosystem complicated. Clean-sheet approach would simplify new systems' design, but it is almost never possible, due to the need of integrating the novel technologies with the legacy ones.

Considering the solutions and the standardization of 5G and beyond, the feature set is expected to further evolve over time and the new technology elements will be added, making the system even more complex. Because of that, the features' set will have to be managed in adaptive and automated way, evolving towards self-learning mechanisms [4]. On the other hand, there are certain features and spectrum bands which have different advantages and disadvantages, making them suited for different applications or use cases, and it is required to use combination of them to fulfill the requirements of those applications or use cases in a unified manner. The authors conclude that by taking the above discussion into account, and in order to make an efficient use of different features and considering various scenarios, the radio resources coordination shall be addressed on multiple levels. Additionally, the non-uniform traffic demand, evolves towards more complex cases, along with the introduction of new service types, like vehicular communication or Industrial IoT. To cope with such data demands in future networks, the access node capabilities, their locations and density should be deployed in a non-uniform manner [5]. From the above discussion, the authors conclude that there is a need to provide a scheme to manage functions in a unified manner, in order to minimize the complexity of adding new features, bands, access methods, transmission schemes into the system, and, in turn, to decrease the need for redesign.

*Approaches to Complex Networks Management*

The authors see three possible ways to handle the services with diverging requirements (see *Figure 2*):

- design separated systems to realize different requirements which results in fragmentation of the solutions. Introducing a new system every time when there is a need to meet the requirements of a new application and integrating with the existing systems is very costly;
- design a single system to meet all the necessary requirements. The drawback of this approach is that it is essential to meet all the requirements in advance and make an over-dimensioned and complex system;
- providing a natively unified and hierarchical approach to the design of the system, with abstraction layer(s) in order to easily introduce new features and meet the requirements not known in advance. The basic assumption is to design optimized solutions on the lower layer and generic, unified mechanism on top.

The authors propose to handle the heterogeneity of Radio Access Technologies (RATs), spectrum bands and types, devices, service mixes and features by hybrid framework with three main components: a *unified upper layer* (handling the context independently of the underlying technology), an *abstraction middle layer* (enabling an "easy" add-on of the techniques below and making the upper layer independent of the specifics of the specialized solutions) and a *specialized lower layer* (to best serve a particular purpose). This model as such is already present in the literature. Some aspects of holistic RRM have been addressed in METIS-II project [6]. The application of abstraction layers within RAN was discussed in COHERENT project [7], while a good example with detailed API description is provided in [8]. What the authors propose in this paper and what is described in the next sections is to use the approach as a native design for the current and future systems throughout the whole RRM architecture stack (including "low-level RRM", "high-level RRM" and network management and orchestration layers).

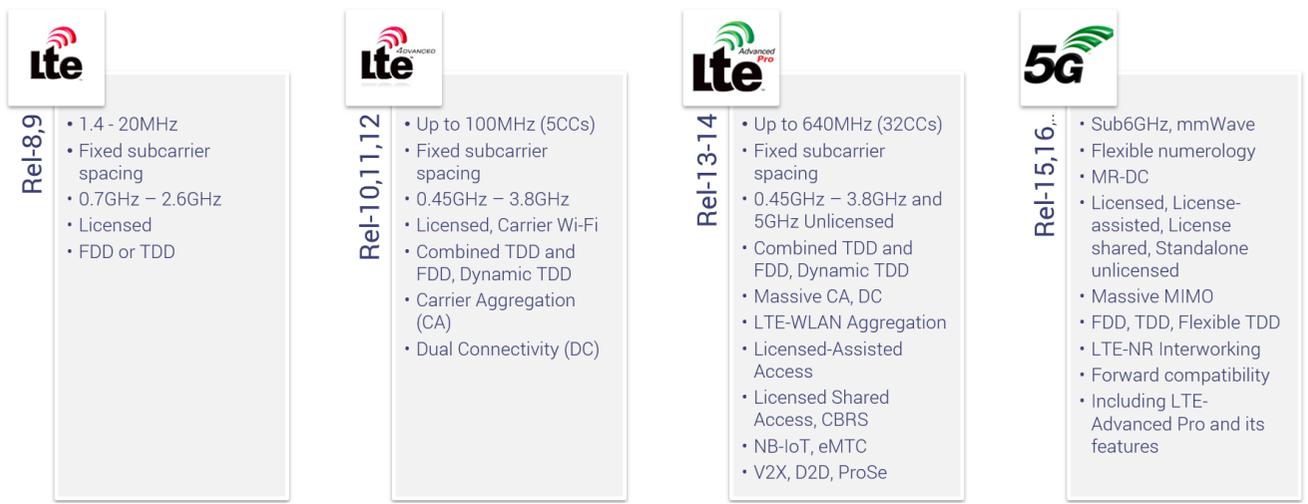

*Figure 1. LTE-to-5G systems' evolution*

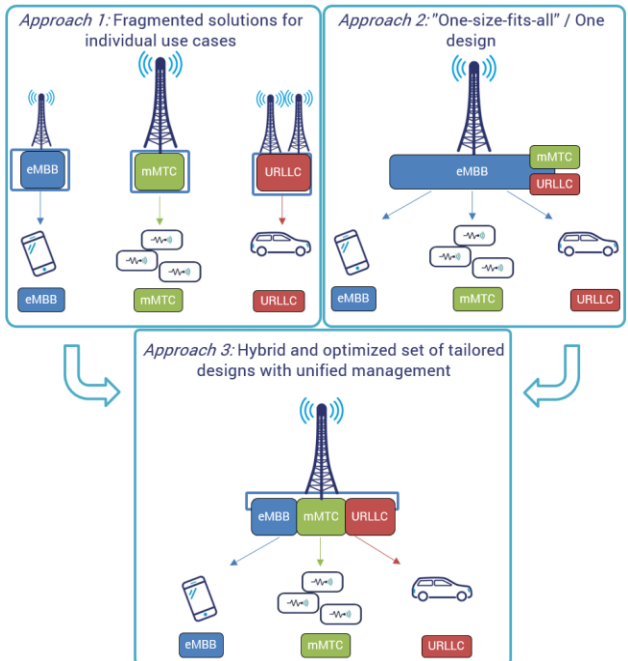

*Figure 2. Three design approaches for management of complex networks*

## III. Resource Management Framework

The authors propose to create the framework using the following actions: *encapsulate* (the solution), *simplify* (the solution), *hide* (the solution), *expose* (the simplified version of the solution), and *use* a coordination scheme. As an example, instead of proposing several specific options with different names (like in LTE-NR tight RAN integration options) or specific band combinations (e.g. for CA application), provide the description of properties and capabilities. This requires to create abstraction layers on multiple levels, which allows to separate different aspects and plug-in new features, depending on where they fit optimally. The proposed framework is based on three design principles: hierarchization, specialization and abstraction.

### Hierarchization

Hierarchy of the RRM and orchestration enable the creation of independency of the unified upper layer from the lower-layer tailored solutions. The examples in 5G context include hardware-software (HW/SW) split, Control Plane/User Plane (CP/UP) split, Self-Organizing Networks (SON) over RRM, separate RAN and CN evolutions. The basic idea of the hierarchy in resource management is the following: firstly, select traffic types that need to be served (or spectrum licensing schemes, or radio technologies to be used) and then select proper resource management algorithms (e.g. individual schedulers) and let them operate. If we apply this to the MAC layer, the resource management algorithms could represent individual schedulers or access mechanisms managed by a scheduling coordinator. If we apply this to RRC layer, the individual elements could be different cells or RATs managed by a single traffic steering. This approach allows to adjust the policies and operation according to the current traffic situation (e.g. large MTC traffic during a specific part of the day, or large data download sessions in the evening).

### Modularization, Specialization and Optimization

In the context of variety of different aspects in complex systems, the specialization and optimization refer to lower level algorithm or function, which is tailored to service, traffic type, feature, spectrum type etc. An example of such on MAC layer is a tailored mechanism for accessing spectrum, e.g. a specific Random Access mechanism dedicated for "one-shot" transmission of MTC sporadic traffic, proposed by the first author in [9], while a different traffic type can be handled by a regular/other Random Access procedure. Another example on Packet Data Convergence Protocol (PDCP) layer is use of NR cell for handling an URLLC traffic type, while LTE cell for a regular MBB traffic for load balancing purpose, both being available in the same area.

### Independency, Unification and Abstraction Layer

The abstraction layer enables a straightforward addition of new elements and to simplify the upper layer of the hierarchical mechanism, which does not have to know all the specifics. An example, in this context, is the use of the PHY layer representation, which have sophisticated accuracy to capture all the essential characteristics of the PHY layer processes, while, at the same time, not exposing those specifics to the MAC layer. In this design, the MAC layer is not aware of the details, but is only given certain parameters to manipulate with and "knows" which waveform or algorithm to use based on higher level policies or learning from the obtained performance indicators. In this particular example, the abstraction is provided via the model based on SINR calculation, taking into account the specifics of a particular waveform. The MAC layer implementation and the scheduler is independent of the lower level PHY layer. MAC is only aware of "what" use case or traffic type it is suitable for; what are the high-level parameters it can decide to use; what are the metrics to measure the performance. Switching to higher layer management, e.g. in the SON field, the SON coordinator needs input from the underlying SON features (like Mobility Robustness Optimization, MRO or Mobility Load Balancing, MLB) through an abstracted manner, e.g. using a common unit (like dB). Therefore, the abstraction layer serves as a translator from whatever input value to a dB, for the SON coordinator to make a unified decision. In this context, the conclusion from [5] is that it is possible to unify the operation of the various individual RRM features, utilizing the same RRC procedures to be invoked when the action is to be taken. Individual functions (e.g. MRO function) invoke certain actions, based on the same or different measures, which, in turn, trigger the lower level procedures (i.e. X2 or RRC). By creating a unified view on this, it is possible to provide a consistent operation, independent of the scenarios, available functions or traffic patterns [5].

*The Unified and Hierarchical Framework*

Gathering all the principles from previous sections altogether allows to create a hybrid architecture, which is defined as: *an application of tailored resource management mechanisms, targeted to specific, or individual traffic type, or application scenario. It allows to use optimized algorithms to fulfil traffic types' requirements and specifics, and encapsulate them without the need to expose their specifics to the upper layer*. Based on this, the proposed novel generic framework is presented in *Figure 3a)*. It covers three layers: *unified upper layer* (context-aware), *abstraction middle layer* (encapsulating the lower layer solutions) and *specialized lower layer* (mechanisms tailored to specific needs). The unified management layer is "context- or service-aware", and is able to select the lower layer aspects (e.g. the amount of resources, prioritization of an algorithm, or function) based on the requirements and feedback. The abstraction layer is needed to translate the upper-layer parameters to the lower-layer specialized algorithms. Those optimized solutions, laying in the bottom of the stack, are designed to solve particular and very specific problems, that may not be known in detail to the upper layer algorithm. By proper encapsulation, a recursive pattern can be created, where the optimized solution can be further decomposed onto another level of hierarchy, embedding even lower level optimized solutions into a module, which exposes certain characteristics to the higher-level abstraction layer, which, in turn, is managed by higher level unified management.

*Challenges and Costs*

The challenges for the proposed framework include:
- *System design mindset* - changing the way, the systems are designed: modularity, abstraction and coordination shall be incorporated as principles on various layers.
- *Design of the abstraction layer* – this is the most sensitive aspect. The more details related to the lower layer mechanisms are added (allowing more flexibility for control), the more complicated and heavier the abstraction layer is. In contrary, the less details (simplifying the abstraction layer), the lower adjustment possibilities for the lower layer mechanisms control.
- *Verifying the cost vs performance* – is not straightforward and cannot be generalized, as depends on the layer on which the framework is applied and on the types of the mechanisms under control.
- *Using coordination as a principle* – requires to provide interfaces between the entities that are subject to be centrally coordinated;
- *Signaling overhead* – relates to coordination signaling resulting from the interworking between coordination and controlled entities. The interaction includes inputs and actions taken by the coordinator, and feedback from the mechanisms under control. The amount of signaling depends on the layer on which the framework is applied.

## IV. Application of the Framework

*Unified MAC Layer*

The framework application at the MAC level allows to define the architecture, where there is a general MAC scheduler, serving as a coordinator of individual "lower layer schedulers" (or access mechanisms). This, in turn, enables to design or select a proper and optimal solution for particular need, while the generic and unified mechanism takes care of proper communication between internal schedulers. The individual mechanisms, can be e.g. [9]:
- dynamic, channel adaptive resource scheduling for high-volume broadband traffic using standard resource scheduling mechanisms;
- semi-static-scheduling for traffic requiring relaxed synchronicity. MAC coordinator decides on the amount of resources allocated for this type of traffic, since the scheduler will not adapt to specific channel conditions;
- one-shot transmission with contention-like based approach, for sporadic MTC traffic, enabling payload transmission in physical layer random access channel. For this scenario, MAC coordinator only defines and steers the amount of resources dedicated for random access channel.

*Figure 3b)* presents the mapping of the generic scheme, to this use case. The Unified MAC coordinates the distribution of the incoming traffic (based on its specifics) to the radio frame. The specific traffic type is handled accordingly, using the dedicated access scheme (e.g. the MTC traffic is handled with the use of one-shot transmission procedure, while the MBB traffic is dynamically assigned resources). The higher-level scheduler simply distributes the available resources to different traffic types, according to the "collective traffic demand" of that traffic type, while the lower level access mechanisms deal with individual transmissions on an instantaneous basis. This approach is different to the operation of the scheduler with respect to the standardized MAC in 5G NR, an example of which is provided in [10]. As per 5G requirements, the scheduler shall support different types of traffic and services with different requirements. It also shall support operation using different subcarrier spacings and limited bandwidth for some devices. However, it is still a single entity having a lot of different options and is provided as a single block taking all parameters into its operation. This implies, that whenever a new type of service or waveform or access mechanism is added to 5G standard, requires redesign of the MAC scheduler to incorporate those changes. What the authors propose here, which is different, is to separate the concerns, where the scheduling mechanisms (and/or PHY layer) are separated from the MAC coordinator, which allocates the resources on the higher "sub-layer". Another application of this approach can be its use for the Dynamic Spectrum Sharing (DSS) feature between NR and LTE [11]. Here, the unified scheduler (upper layer) selects (based on the number of users, device capabilities and traffic types), the radio interface. The design could be as follows: under the lower layer of the hierarchical scheduler, there is another scheduler that knows the specifics of handling LTE-capable and connected users, and another scheduler to handle NR users. The unified upper layer scheduler does not have to know, that there is something called "LTE radio" or "NR radio". Instead, it has to be aware of the properties and capabilities to coordinate the distribution and assigned portions of the traffic, bearers or users to specific radio interface. When there is a new PHY layer (or a waveform, or access method) introduced, the same approach is taken and it does not change the overall design.

## Unified PDCP Flow Control

*Figure 3c)* provides an application of the framework for PDCP flow control mechanism. Unified PDCP covers all the different options for distributing packets to different "connection legs" (i.e. connections, cells, or access points). Examples of those include: LTE DC, NR DC, multi-RAT DC (like LTE-NR DC, NR-LTE DC), or LWA. The DC scheme can also be applied for different purposes, e.g. for improving throughput (i.e., sum of the data from both cells), for fast load balancing (i.e., when all the packets go through only one cell, while two cells are allocated), for improving reliability in URLLC cases (i.e., by using PDCP duplication feature within NR). Therefore, the flow control mechanism can be generic, subject to load, preferences, purpose, while the underlying "legs" can be LTE, NR, Wi-Fi, in different combinations. Whenever a new radio interface shows up within standardization, it is very easy to incorporate it in that framework.

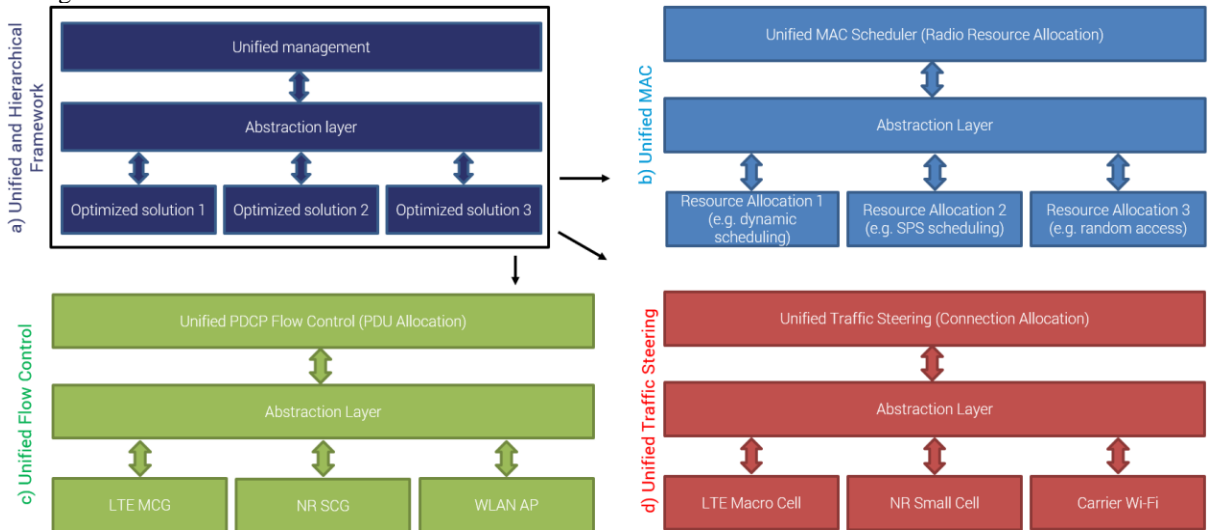

**Figure 3.** *Unified and Hierarchical Framework and Application Examples*

## Unified Traffic Steering (UTS)

The design of the UTS focuses on orchestration of the available spectrum access mechanisms (the simplified scheme is shown in *Figure 3d*). It also enables a scenario-specific feature prioritization and ranking to optimally utilize radio resources. UTS logic serves as an engine, coordinating TS mechanisms. This approach aims at avoiding network instabilities and contradicting actions being taken by individual RRM features when fragmented solutions are implemented. The UTS concept design enables the integration of new features, additional spectrum bands or novel spectrum licensing schemes, as investigated in [5]. In this framework, the intermediate layer, being the unified function, allows to use multiple higher-level algorithms, managing the lower level functions' triggering. The common layer is RRC protocol, based on which, the actions have impact on the particular user situation. The framework design enables flexible adaptation to the particular scenario and the availability of features in a specific place. The independency, provided by unification, is an elegant way to incorporate new functions, not known at the time when the framework is created – and thus provides flexibility. The adaptive UTS framework assumes awareness of the traffic demand and the ability to optimize its power consumption within HetNets. A simple and somewhat obvious example can be an application of high and low frequency carriers and their combination with small cells and a DC feature in dense urban area, while using only low frequency macro cells on highways, or switching off small cells/high frequency layer during night in the business centres during night.

New features (e.g. SON/RRM algorithm) or new managed objects (e.g. frequency bands), are provided with a unified procedure, allowing them to be added to the framework. Based on this design, the individual feature can be incorporated "above" or "below" UTS logic. What needs to be done is to define: *what are the inputs and outputs* (measures and actions), *where does the item belong to* (RRM-low level, RRM-high level, SON, radio interface), *how does it interact with other functions* and *what scenario does it fit*. An example of introduction of a new feature into the framework is as follows. Let's consider a CA feature, going through the above questions one by one:

- *Location in the framework:* "below" UTS, i.e. UTS controls the use of CA; CA relates to connected mode mobility;
- *Inputs to CA from UTS:* Primary cell allocation, Secondary cell(s) add/release of particular Component Carrier for a particular UE;
- *CA interacts with:* LAA (CC can be in unlicensed spectrum), DC (secondary cell may use CA and/or DC may be used instead of DC for capacity improvements);

- *Application scenarios:* macro-networks (each site may use multiple CCs), HetNets (Secondary eNB may use multiple CCs).

The abstraction layers can be defined at multiple points (see *Figure 4* for example design taking those into account):
- the UTS logic input is taking actions from multiple features;
- the MNO strategies are independent of the actual features and are based on common parameter being the network load which is technologically independent;
- the overall operation is independent on the actual and specific network layers and frequency bands or carrier frequencies;
- the configuration of the switching points between the individual features' operation can be MNO- and scenario-specific and is independent of the set of available features.

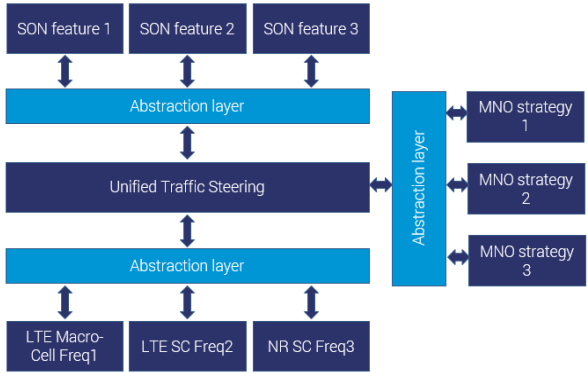

*Figure 4. Unified Traffic Steering Interfaces*

### Joint Architecture for All Layers

The extension of the proposed framework is to merge the Unified MAC design with Unified PDCP and UTS. The selection of a particular RAT (e.g. LTE or NR) can be based on the common set of parameters. Once the RAT is selected, the proper scheduler can be in operation, and feedback can be provided to adapt the upper layer UTS. Another option is that the UTS can decide to use several nodes with different RATs (e.g. Master Node, MN with LTE and Secondary Node, SN with NR) for one user and another configuration (e.g. Master on NR and Secondary on LTE) for another user. In this context, the unified scheduler needs to coordinate the resources used for both users, without looking on the specifics of each radio interface. Unified MAC is focusing on a set of users (MAC layer provides access to multiple users in a single cell), while UTS is focusing on a specific user (decides which cells to assign to a particular user). Unified MAC operates on various lower layer PHY schemes, while UTS uses a common set of RRC procedures and operates on a different set of RATs and features. The joint architecture is provided in *Figure 5* where all individual elements are combined with several abstraction layers in-between.

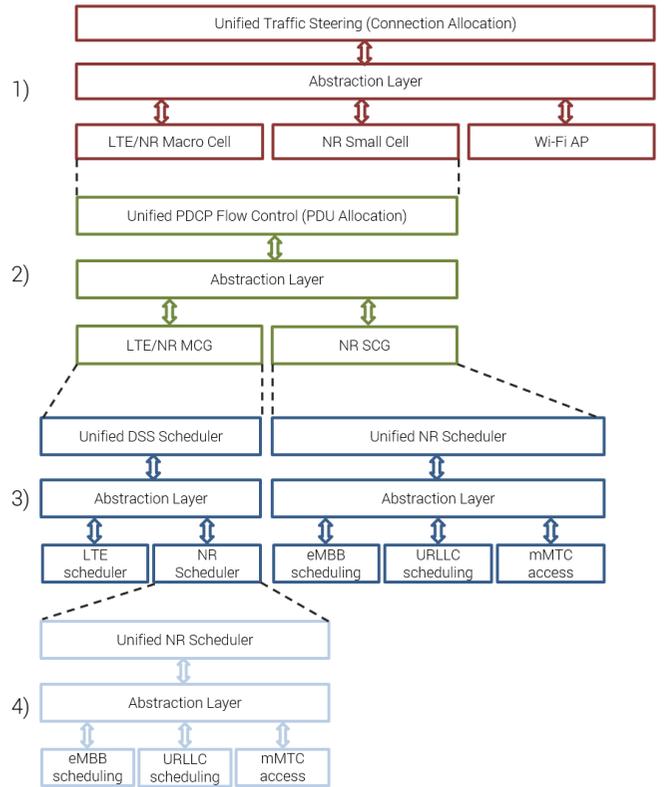

*Figure 5. Joint Approach for Hierarchical and Modular RRM Architecture*

The operation of the configuration from *Figure 5* is as follows:
1) UTS decides on an assignment of the resources to particular user, from the available ones, including combined LTE/NR macro-cell (i.e. a cell with DSS feature enabled), NR small cell and Wi-Fi AP. For this particular example, UTS decided to apply NR-DC, with NR macro-cell as MN and NR SC as SN. The resources are allocated and the data is going through split bearer to the PDCP at MN;
2) PDCP flow control mechanism distributes individual packets to the "legs" allocated by UTS based on the load and other factors;
3) The data from both "legs" are allocated to PHY resources using MAC schedulers. In case of SN, MAC coordinator is distributing resources among users using schemes related to 5G services, in our case, the service is MBB, thus, the selected algorithm is allocating resources based on that principle, where the MAC coordinator is simply gathering the overall traffic for eMBB, mMTC and URLLC and allocating specific portion of resources, and the lower layer scheduler allocates resources within "eMBB area".
4) There is a different situation within MN, which is subject to DSS feature, i.e. it serves both LTE and NR users. In this case, there is a recursive pattern of our proposed scheme implemented. Namely, the most upper coordinator distributes PHY resources between LTE and NR, while the NR users are served by NR MAC coordinator, which in turn triggers the lowest level specific schedulers as in 3).

## V. Further Generalization of the Approach

One direction of future work is to generalize this approach further and extend to higher layers, including transport network, core network, IP layer, and the service layer, where aspects, like load balancing and traffic steering, play important role as well. The examples of "SON for EPC" are presented in [12] and include: MME load balancing to manage connections between the signaling nodes, tracking area list optimization, or routing optimization for latency decrease between Serving Gateways. The examples for "SON for services" include [12]: end-to-end voice quality improvements, and IP Multimedia Subsystem (IMS) nodes' load balancing. The Software-Defined Network (SDN)-based systems abstract the data-plane from the management and control, while Network Functions Virtualization (NFV) concept abstracts the network functions from the underlying infrastructure, which suits the general idea of the proposed unified framework. In this context, "softwarization" of the core networks is already present in 5G context, by means of Service-Based Architecture, defining the abstraction and common communication platform using REST API.

Another aspect from the 5G scope, to be further studied in the context of hierarchical management framework, is network slicing and its application in NG-RAN. Recent works on this topic included virtualized RAN and dynamic RAN slicing, and can be found in e.g., [13][14]. In the context of slicing, three areas in the RAN management could be defined (see *Figure 6*), namely: *intra-slice management* – "slice-specific" (optimization of resources for a certain slice – i.e. all users being connected to a given slice), *inter-slice management* – "slice-unified" (e.g. MAC scheduler providing access resources to multiple slices and needing to distribute them among slices), and *slice-independent management* – "slice-independent" (e.g. traffic steering – focusing on a certain user, which can be connected to multiple slices at once). ITU is also working on the unified logical architecture for Machine Learning applications [15], where the proposed scheme can be applied.

## VI. Summary

Concluding the aspects discussed in this paper, it can be seen that the management of the network resources, within the complexity visible in today's mobile networks and ongoing developments, should be addressed in a generic manner, to be able to operate those networks efficiently. The development does not stop, while *backwards compatibility* is one of the requirements in practically implemented systems. It is not possible to design a new system to just cope with the new requirements from scratch. This results in the need of adding new items to the existing system on different levels (e.g., new RAT, new function, new algorithm, new service), which requires adjusting the existing networks. The adaptation of the existing system to the new requirements and functionalities is a complex task, and can be solved by one of the possible approaches: *universal* – where all of the different requirements and services are captured by a single scheme and, by using reconfigurations, they can be optimized; *individual and specific* – each new service gets a new system version – it is costly and creates fragmentation in the market; *unified and hierarchical* – where the specifics are separated from the system's coordination and new items can be added as "plugins" to the architecture. The authors claim that, it is possible to generalize the approach and provide a framework to cope with those challenges on various network layers. The framework aims at simplification of the introduction of new elements. The authors claim that the unification, hybridization and hierarchization can be the approach for future network management, including "RRM-low", "RRM-high", SON and network orchestration.

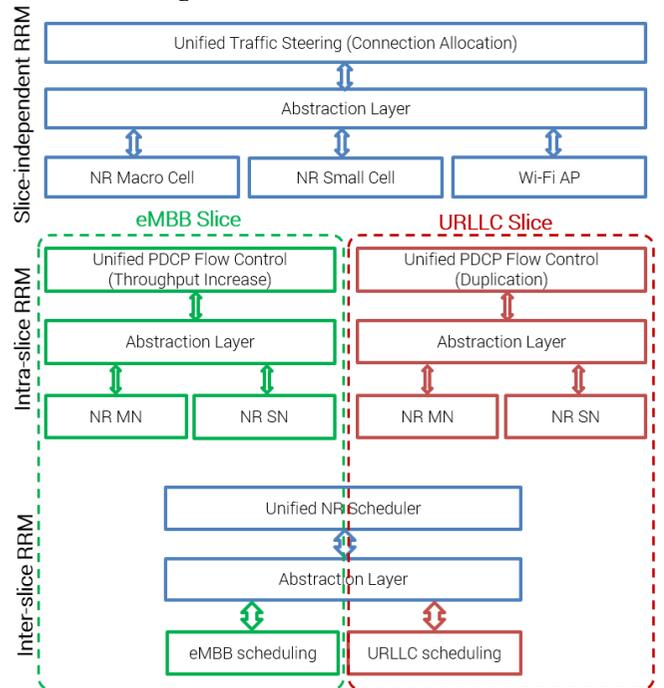

*Figure 6. Hybrid and Modular Framework within Slicing Context*

## Acknowledgements

This paper bases on the PhD dissertation of Marcin Dryjanski.

The work by Adrian Kliks was funded with national subvention for scientific investigations by MNiSW for 2019 (08/81/SBAD/8148).

**Biographies**

*Marcin Dryjanski* (SM) holds Senior IEEE Membership since 2018. He was a Work-package Leader in FP7-5GNOW project. Marcin served as a RAN specialist working on architecture towards 5G. He is a co-author of research papers on 5G design, and a book: "From LTE to LTE-Advanced Pro and 5G" published by Artech House in 2017. Marcin received Ph.D. with honours from Poznan University of Technology in Poland in 2019.

*Adrian Kliks* (SM) is an assistant professor at Poznan University of Technology's Department of Wireless Communications, Poland. His research interests include new waveforms for wireless systems applying either non-orthogonal or non-contiguous multicarrier schemes, cognitive radio, advanced spectrum management, deployment and resource management in small cells, and network virtualization.